\documentclass[twocolumn,superscriptaddress,nofootinbib]{revtex4-2}
\usepackage{mathtools,amssymb,bm,graphicx,hyperref}
\allowdisplaybreaks[1]

\renewcommand\d{\partial}
\newcommand\<{\langle}
\renewcommand\>{\rangle}
\newcommand\0{{\bm{0}}}
\renewcommand\k{{\bm{k}}}
\newcommand\p{{\bm{p}}}
\renewcommand\r{{\bm{r}}}
\newcommand\D{\mathcal{D}}
\renewcommand\O{\mathcal{O}}
\newcommand\Z{\mathbb{Z}}

\begin{document}

\title{Efimov effect at the Kardar-Parisi-Zhang roughening transition}

\author{Yu Nakayama}
\affiliation{Department of Physics, Rikkyo University,
Toshima, Tokyo 171-8501, Japan}
\author{Yusuke Nishida}
\affiliation{Department of Physics, Tokyo Institute of Technology,
Ookayama, Meguro, Tokyo 152-8551, Japan}

\date{October 2020}

\begin{abstract}
Surface growth governed by the Kardar-Parisi-Zhang (KPZ) equation in dimensions higher than two undergoes a roughening transition from smooth to rough phases with increasing the nonlinearity.
It is also known that the KPZ equation can be mapped onto quantum mechanics of attractive bosons with a contact interaction, where the roughening transition corresponds to a binding transition of two bosons with increasing the attraction.
Such critical bosons in three dimensions actually exhibit the Efimov effect, where a three-boson coupling turns out to be relevant under the renormalization group so as to break the scale invariance down to a discrete one.
On the basis of these facts linking the two distinct subjects in physics, we predict that the KPZ roughening transition in three dimensions shows either the discrete scale invariance or no intrinsic scale invariance.
\end{abstract}

\maketitle

\section{Introduction}
The Kardar-Parisi-Zhang (KPZ) equation for surface growth~\cite{Kardar:1986},
\begin{align}\label{eq:KPZ}
\frac{\d h}{\d t} = \nu\nabla^2h + \frac\lambda2(\nabla h)^2 + \sqrt{D}\,\eta,
\end{align}
has been a paradigmatic model in nonequilibrium statistical physics~\cite{Barabasi:1995,Lassig:1998,Quastel:2015,Sasamoto:2016,Takeuchi:2018}.
Here, $h=h(t,\r)$ represents a height of $d$-dimensional surface, which grows under a white Gaussian noise obeying $\<\eta(t,\r)\>=0$ and
\begin{align}\label{eq:noise}
\<\eta(t,\r)\eta(t',\r')\> = \delta(t-t')\delta(\r-\r').
\end{align}
The roughness of surface is characterized by the asymptotic scaling of the height-difference correlation function,
\begin{align}\label{eq:scaling}
\<[h(t,\r)-h(0,\0)]^2\> \sim r^{2\chi}\,F\!\left(\frac{t}{r^z}\right),
\end{align}
where $\chi$ is the roughness exponent and $z$ is the dynamical exponent.
The surface is said to be smooth for $\chi<0$ and rough for $\chi>0$.
In one dimension, these scaling exponents can be determined exactly to find that the surface is always rough with $\chi=1/2$ and $z=3/2$~\cite{Kardar:1986}, which has also been confirmed experimentally~\cite{Takeuchi:2014}.

\begin{figure}[b]
\includegraphics[width=0.9\columnwidth]{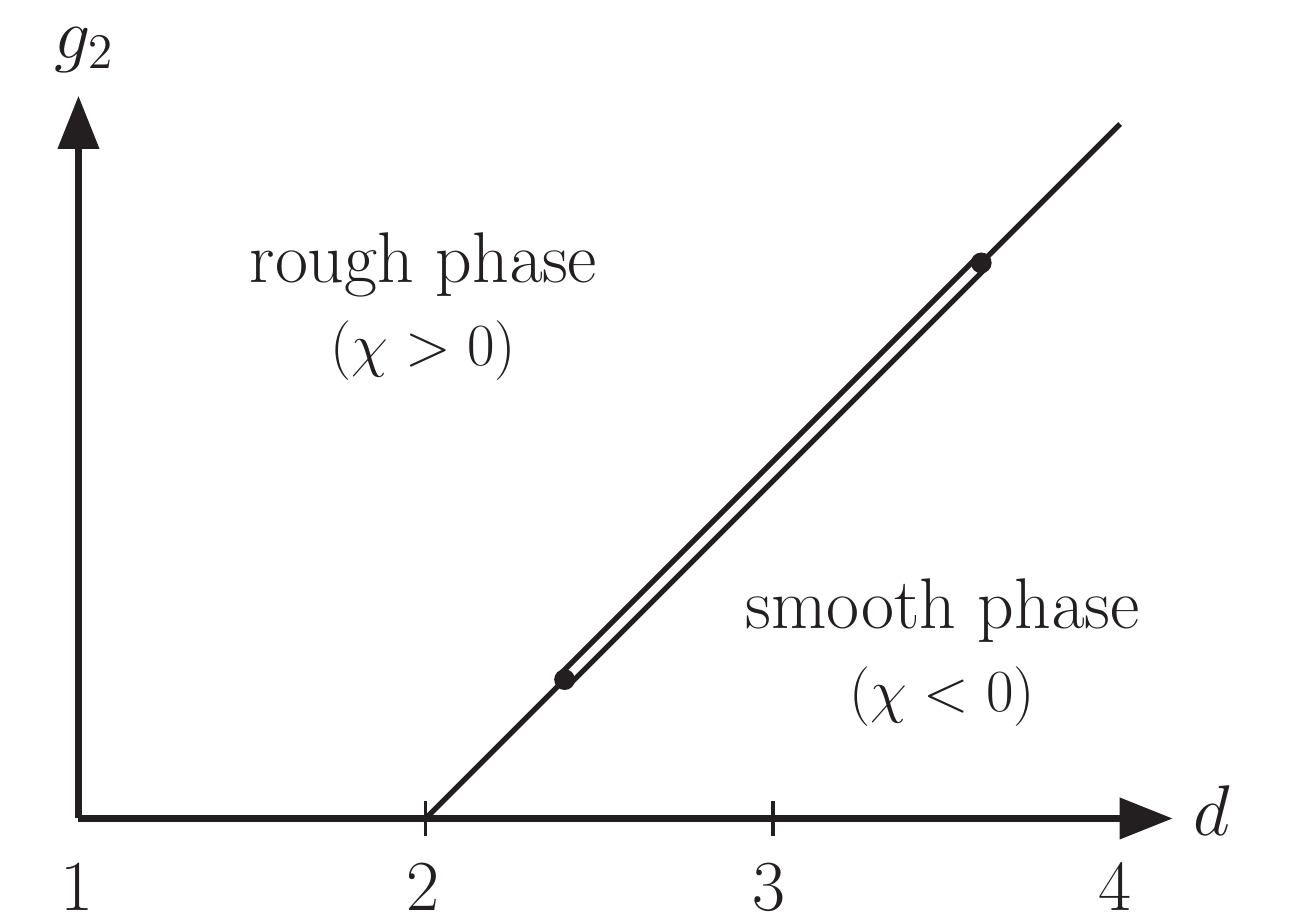}
\caption{\label{fig:phase_diagram}
Schematic phase diagram in the plane of the dimensionality $d$ and the nonlinearity $g_2=D\lambda^2/(4\nu^3)$.
It consists of smooth ($\chi=1-d/2<0$, $z=2$) and rough ($\chi>0$, $z=2-\chi$) phases separated by the roughening transition at which $\chi=0$ and $z=2$ (single line).
It is to be predicted in this paper that there is a finite interval including $d=3$ but not $d=2,\,4$ (double line) where the roughening transition shows either the discrete scale invariance or no intrinsic scale invariance.}
\end{figure}

Physics in higher dimensions is even richer.
While the nonlinear term proportional to $\lambda$ in Eq.~(\ref{eq:KPZ}) is relevant below two dimensions, it turns irrelevant above two dimensions~\cite{Kardar:1986}.
Therefore, the surface growth for a sufficiently small $\lambda<\lambda^*$ is governed by the linear Edwards-Wilkinson equation~\cite{Edwards:1982}, which finds the surface to be smooth with $\chi=1-d/2<0$ and $z=2$ (smooth phase in Fig.~\ref{fig:phase_diagram}).
However, with increasing the nonlinearity, there exists a phase transition at $\lambda^*$, called the roughening transition, at which the surface is marginally rough with $\chi=0$ and $z=2$~\cite{Tang:1990,Nattermann:1992,Doty:1992}.
Then, the surface turns rough for a larger $\lambda>\lambda^*$ (rough phase in Fig.~\ref{fig:phase_diagram}), where the scaling exponents were numerically estimated in Ref.~\cite{Odor:2010} among others at $\chi\approx0.395(5)$ for $d=2$, $\chi\approx0.29(1)$ for $d=3$, $\chi\approx0.245(5)$ for $d=4$, and $\chi\approx0.22(1)$ for $d=5$, with $z=2-\chi$ imposed by the ``Galilean'' invariance~\cite{Medina:1989}.
On the other hand, there have been a number of claims that $d=4$ is an upper critical dimension beyond which the surface is only marginally rough with $\chi=0$~\cite{Halpin-Healy:1990,Bouchaud:1993,Doherty:1994,Blum:1995,Moore:1995,Lassig:1995,Bundschuh:1996,Lassig:1997,Bhattacharjee:1998,Colaiori:2001,Katzav:2002,Fogedby:2005,Fogedby:2006}, although it contradicts numerical simulations of models belonging to the KPZ universality class~\cite{Ala-Nissila:1993,Ala-Nissila:1998,Kim:1998,Castellano:1998,Kim:1999,Marinari:2000,Marinari:2002,Perlsman:2006,Schwartz:2012,Pagnani:2013,Kim:2013,Kim:2014,Alves:2014,Kim:2015}.
The very existence of the upper critical dimension has been one of the most controversial issues regarding the KPZ equation.

What we shall argue in this paper is that already in three dimensions some peculiarity possibly emerges at the roughening transition.
Our argument is based on the facts that the KPZ equation can be mapped onto quantum mechanics of attractive bosons with a contact interaction~\cite{Kardar:1987} and that such three bosons exhibit the Efimov effect in three dimensions so as to break the scale invariance down to a discrete one~\cite{Efimov:1970,Efimov:1973}.
Although each of them is well known in each community, it may not in the other community.
Therefore, we shall first describe these subjects in a brief but self-contained manner in Secs.~\ref{sec:KPZ} and \ref{sec:RG}, respectively, and then draw our conclusion in Sec.~\ref{sec:conclusion}.

\section{From KPZ to attractive bosons}\label{sec:KPZ}
Let us first regularize the white noise obeying Eq.~(\ref{eq:noise}) by replacing it with a colored noise,
\begin{align}
\<\eta(t,\r)\eta(t',\r')\>_\Lambda = \delta(t-t')V_\Lambda(\r-\r'),
\end{align}
where $V_\Lambda(\r)$ is assumed to have a finite range set by $1/\Lambda$ with $\lim_{\Lambda\to\infty}V_\Lambda(\r)=\delta(\r)$, such as $V_\Lambda(\r)=(\Lambda/\sqrt\pi)^d\,e^{-(\Lambda\r)^2}$.
A solution to Eq.~(\ref{eq:KPZ}) for a given configuration of $\eta$ is denoted by $h_\eta$ and we are interested in its ensemble average over the Gaussian noise,
\begin{align}\label{eq:average}
& \<\O(h)\>_\Lambda \equiv \int\!\D\eta\,\O(h_\eta) \notag\\
&\quad \times \exp\!\left[-\frac12\int\!dtd\r d\r'
\eta(t,\r)V_\Lambda^{-1}(\r-\r')\eta(t,\r')\right].
\end{align}
Formally, $h_\eta$ can be expressed by
\begin{align}
\O(h_\eta) &= \int\!\D h\,\O(h) \notag\\
& \times \prod_{t,\r}\delta\!\left[\frac{\d h}{\d t}
- \nu\nabla^2h - \frac\lambda2(\nabla h)^2 - \sqrt{D}\,\eta\right],
\end{align}
where the $\delta$ functions single out $h$ solving Eq.~(\ref{eq:KPZ}) and the Jacobian determinant is unity if the time derivative is understood as It\^o's forward differential operator~\cite{Cardy:1999}.
By exponentiating the $\delta$ functions at the expense of introducing an auxiliary field $\bar{h}$,
\begin{align}
& \O(h_\eta) = \int\!\D h\D\bar{h}\,\O(h) \notag\\
& \times \exp\!\left[-\int\!dtd\r\,i\bar{h}\left\{\frac{\d h}{\d t}
- \nu\nabla^2h - \frac\lambda2(\nabla h)^2 - \sqrt{D}\,\eta\right\}\right],
\end{align}
the functional integration over $\eta$ in Eq.~(\ref{eq:average}) can be performed to obtain the path integral representation of the correlation function,
\begin{align}
\<\O(h)\>_\Lambda &= \int\!\D h\D\bar{h}\,\O(h)\,e^{-S_\Lambda[h,\bar{h}]},
\end{align}
with the action provided by
\begin{align}\label{eq:action}
S_\Lambda[h,\bar{h}] &= \int\!dtd\r\,i\bar{h}\left[\frac{\d h}{\d t}
- \nu\nabla^2h - \frac\lambda2(\nabla h)^2\right] \notag\\
& - \frac{D}{2}\int\!dtd\r d\r'i\bar{h}(t,\r)V_\Lambda(\r-\r')i\bar{h}(t,\r').
\end{align}
This is the field theoretical formulation of stochastic differential equations \`a la Martin, Siggia, Rose, Janssen, and De~Dominicis~\cite{Martin:1973,Janssen:1976,DeDominicis:1976}.

We then apply the Cole-Hopf transformation,
\begin{align}
h(t,\r) = \frac{2\nu}{\lambda}\ln\phi(t,\r), \quad
i\bar{h}(t,\r) = \frac\lambda{2\nu}\bar\phi(t,\r)\phi(t,\r),
\end{align}
which eliminates the nonlinear term in Eq.~(\ref{eq:action}) and leads to
\begin{align}
& S_\Lambda[\phi,\bar\phi] = \int\!d\tau d\r\,
\bar\phi(\tau,\r)\left(\frac\d{\d\tau} - \frac{\nabla^2}{2}\right)\phi(\tau,\r) \notag\\
& - \frac{g_2}{4}\int\!d\tau d\r d\r'
\bar\phi(\tau,\r)\phi(\tau,\r)V_\Lambda(\r-\r')\bar\phi(\tau,\r')\phi(\tau,\r').
\end{align}
Here, $\tau\equiv 2\nu t$ and $g_2\equiv D\lambda^2/(4\nu^3)>0$ are introduced so that the resulting action becomes identical to the imaginary-time action for attractive bosons interacting with a finite-range potential.
Therefore, we find that the role of an attraction between bosons is played by the nonlinearity of the KPZ equation, $g_2$, which is adopted as the vertical axis of Fig.~\ref{fig:phase_diagram}.

In order to recover the original KPZ equation with the white noise, we take the limit of $\Lambda\to\infty$, where the above action is reduced into the local form of
\begin{align}\label{eq:boson}
S_\Lambda[\phi,\bar\phi] &\to \int\!d\tau d\r
\bigg[\bar\phi\left(\frac\d{\d\tau} - \frac{\nabla^2}{2}\right)\phi \notag\\
&\quad - \frac{g_2}{4}(\bar\phi\phi)^2 - \frac{g_3}{36}(\bar\phi\phi)^3 + \cdots\bigg].
\end{align}
Naively, only the two-boson contact interaction proportional to $g_2$ should be present, but higher-order terms allowed by symmetries may be generated by integrating out short-distance degrees of freedom.
Such higher-order terms are usually neglected by considering them to be irrelevant under the renormalization group (RG).
However, the three-boson contact interaction proportional to $g_3$ actually turns out to be relevant in three dimensions, as we shall show below.
We note that the first quantized form of attractive bosons also follows from the KPZ equation with the replica method, where the number of replicas corresponds to that of bosons~\cite{Kardar:1987,Calabrese:2010,Dotsenko:2010}.

\section{Renormalization group analysis}\label{sec:RG}
The renormalizations of $g_2$ and $g_3$ can be performed exactly to all orders in their perturbations~\cite{Braaten:2006}.
For later use, the boson propagator from Eq.~(\ref{eq:boson}) in the Fourier space is denoted by
\begin{align}\label{eq:propagator}
G(K) = \frac{-1}{ik_0-\k^2/2},
\end{align}
where $K=(k_0,\k)$ is a set of frequency and wave vector.

\subsection{Two-boson coupling}
\begin{figure}[t]
\includegraphics[width=\columnwidth]{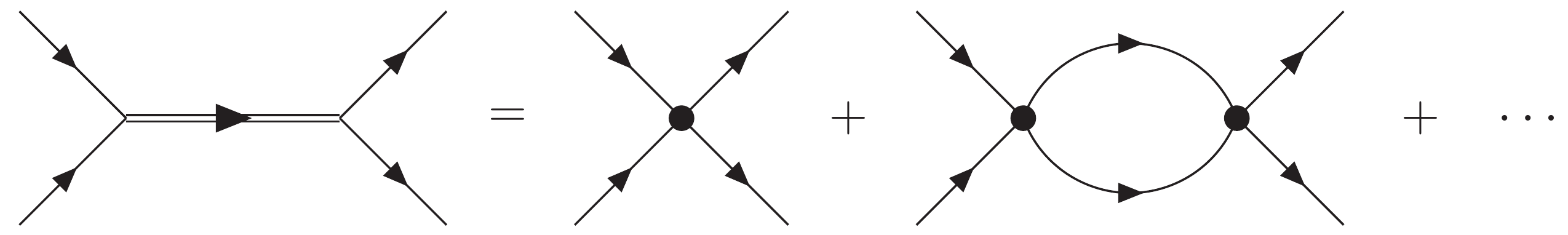}
\caption{\label{fig:2-boson}
Diagrammatic equation for the two-boson scattering amplitude.
The single line represents the boson propagator in Eq.~(\ref{eq:propagator}), the dot the two-boson coupling $g_2$, and the double line the scattering amplitude in Eq.~(\ref{eq:2-boson}).}
\end{figure}

We first study the renormalization of the two-boson coupling $g_2$.
The two-boson scattering amplitude $T_2(K)$ with center-of-mass $K$ is obtained by summing up a geometric series of diagrams depicted in Fig.~\ref{fig:2-boson},
\begin{align}\label{eq:2-boson}
T_2(K) = \left[\frac1{g_2}
- \int\!\frac{d^{d+1}P}{(2\pi)^{d+1}}\frac{G(K/2+P)G(K/2-P)}2\right]^{-1}.
\end{align}
After performing the integration over $p_0$ with the residue theorem, the integration over $\p$ under a sharp-cutoff regularization $|\p|<\Lambda$ leads to
\begin{align}\label{eq:amplitude}
& \frac1{T_2(K)} = \frac{\Lambda^{d-2}}{\hat{g}_2}
- \frac{\Lambda^{d-2}}{(d-2)(4\pi)^{d/2}\Gamma(d/2)} \notag\\
&\quad - \frac{\Gamma(1-d/2)}{2(4\pi)^{d/2}}\left(\frac{k^2}{4}-ik_0\right)^{d/2-1}
+ O\!\left(\frac{k^2}{\Lambda^{4-d}}\right).
\end{align}
Here, a dimensionless coupling $\hat{g}_2\equiv\Lambda^{d-2}g_2$ is introduced.

We now determine the cutoff dependence of $\hat{g}_2$ by requiring the scattering amplitude to be cutoff independent at low frequency and wave vector for $d<4$, where $O(k^2/\Lambda^{4-d})$ in Eq.~(\ref{eq:amplitude}) is negligible.
From the Callan-Symanzik equation,
\begin{align}
\left(\frac\d{\d\Lambda} + \frac{\d\hat{g}_2}{\d\Lambda}\frac\d{\d\hat{g}_2}\right)T_2(K) = 0,
\end{align}
the $\beta$ function immediately reads
\begin{align}
\Lambda\frac{\d\hat{g}_2}{\d\Lambda}
= (d-2)\hat{g}_2 - \frac{\hat{g}_2^2}{(4\pi)^{d/2}\Gamma(d/2)},
\end{align}
where two fixed points are found.
One is the infrared (IR) fixed point at $\hat{g}_2=0$ simply describing free bosons and the other is the ultraviolet (UV) fixed point at $\hat{g}_2^*=(d-2)(4\pi)^{d/2}\Gamma(d/2)>0$ for $d>2$.
The latter describes critical bosons, where a bound state of two bosons is formed with zero binding energy.
Therefore, there exists a transition from unbound ($\hat{g}_2<\hat{g}_2^*$) to bound ($\hat{g}_2>\hat{g}_2^*$) bosons with increasing the attraction.
It is this binding transition of two bosons for $d>2$ that corresponds to the roughening transition (see Fig.~\ref{fig:phase_diagram}).
The above RG equation is indeed consistent with that obtained by the dynamical RG method applied to the KPZ equation~\cite{Tang:1990,Nattermann:1992}.

The two-boson scattering amplitude at the UV fixed point has the scaling form of
\begin{align}
T_2^*(K) = -\frac{2(4\pi)^{d/2}}{{\Gamma(1-d/2)}}\left(\frac{k^2}{4}-ik_0\right)^{1-d/2},
\end{align}
which is scale invariant under the dynamical exponent of $z=2$.
Unless this scale invariance is broken by the higher-order terms in Eq.~(\ref{eq:boson}), such critical bosons also enjoy the nonrelativistic conformal invariance~\cite{Hagen:1972,Niederer:1972,Mehen:2000,Son:2006,Nishida:2007}.

\begin{figure}[t]
\includegraphics[width=\columnwidth]{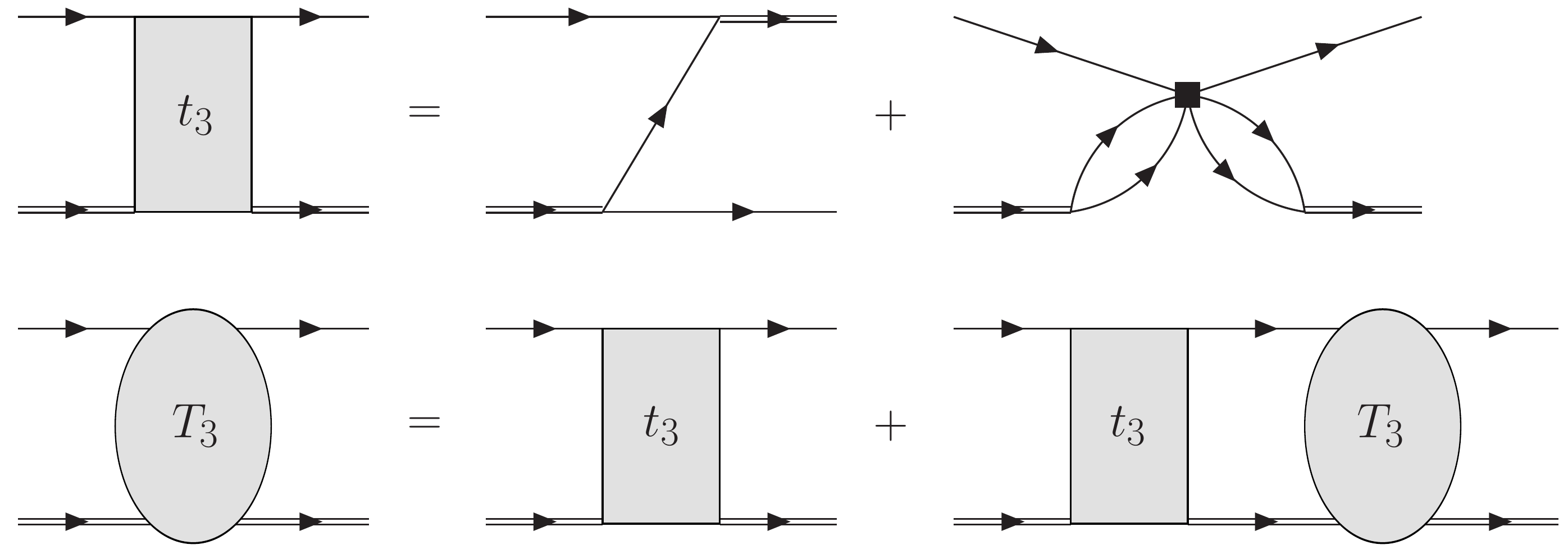}
\caption{\label{fig:3-boson}
Diagrammatic equations for the three-boson scattering amplitudes.
The single and double lines represent the same as in Fig.~\ref{fig:2-boson}, the square the three-boson coupling $g_3$, and $t_3$ and $T_3$ are the tree and full scattering amplitudes in Eqs.~(\ref{eq:tree}) and (\ref{eq:full}), respectively.}
\end{figure}

\subsection{Three-boson coupling}
We then turn to the renormalization of the three-boson coupling $g_3$ at the binding transition $g_2=\hat{g}_2^*/\Lambda^{d-2}$.
The three-boson scattering amplitude between a boson and a pair of bosons from their initial $(K/3+P,2K/3-P)$ to their final $(K/3+P',2K/3-P')$ is denoted by $T_3(K;P,P')$.
Such a full scattering amplitude solves an integral equation depicted in Fig.~\ref{fig:3-boson},
\begin{align}\label{eq:full}
& T_3(K;P,P') = t_3(K;P,P') + \int\!\frac{d^{d+1}P''}{(2\pi)^{d+1}}\,t_3(K;P,P'') \notag\\
&\quad \times G(K/3+P'')T_2^*(2K/3-P'')T_3(K;P'',P'),
\end{align}
with the tree scattering amplitude provided by
\begin{align}\label{eq:tree}
t_3(K;P,P') = G(K/3-P-P') + \frac{g_3}{9g_2^2},
\end{align}
where $\lim_{\Lambda\to\infty}g_2\int d^{d+1}P/(2\pi)^{d+1}[G(K/2+P)G(K/2-P)]/2=1$ following from Eqs.~(\ref{eq:2-boson}) and (\ref{eq:amplitude}) is applied.
Because nonanalyticity in the lower-half plane of $p_0''$ arises only from a pole of $G(K/3+P'')$, the integration over $p_0''$ can be performed with the residue theorem, which sets $ip_0''=(\k/3+\p'')^2/2-ik_0/3$.
Similarly, by setting external $ip_0=(\k/3+\p)^2/2-ik_0/3$ and $ip_0'=(\k/3+\p')^2/2-ik_0/3$, Eq.~(\ref{eq:full}) is reduced into an integral equation solved by the on-shell scattering amplitude $T_3(K;\p,\p')\equiv T_3(K;P,P')|_{ip_0^{(\prime)}=(\k/3+\p^{(\prime)})^2/2-ik_0/3}$.
Finally, the projection onto the $s$-wave component,
\begin{align}
& T_3^{(0)}(K;p,p') \notag\\
&\equiv \frac{\Gamma(d/2)}{\sqrt\pi\,\Gamma(d/2-1/2)}
\int_0^\pi\!d\theta'\sin^{d-2}\theta'\,T_3(K;\p,\p'),
\end{align}
brings the integral equation into
\begin{align}
& T_3^{(0)}(K;p,p') \notag\\
&= t_3^{(0)}(K;p,p') - \frac{4\sin(\pi d/2)}{\pi}
\int_0^\Lambda\!dp''p''^{d-1}\,t_3^{(0)}(K;p,p'') \notag\\
&\qquad \times \left(\frac{3p''^2}{4}+\frac{k^2}{6}-ik_0\right)^{1-d/2}
T_3^{(0)}(K;p'',p'),
\end{align}
under a sharp-cutoff regularization $|\p''|<\Lambda$ with
\begin{align}
t_3^{(0)}(K;p,p') = \frac{{}_2F_1\!\left(\frac12,1;\frac{d}{2};\frac{p^2p'^2}
{(p^2+p'^2+k^2/6-ik_0)^2}\right)}{p^2+p'^2+k^2/6-ik_0} + \frac{\hat{g}_3}{\Lambda^2}.
\end{align}
Here, a dimensionless coupling $\hat{g}_3\equiv\Lambda^2g_3/(9g_2^2)$ is introduced.

\begin{figure}[t]
\includegraphics[width=0.95\columnwidth]{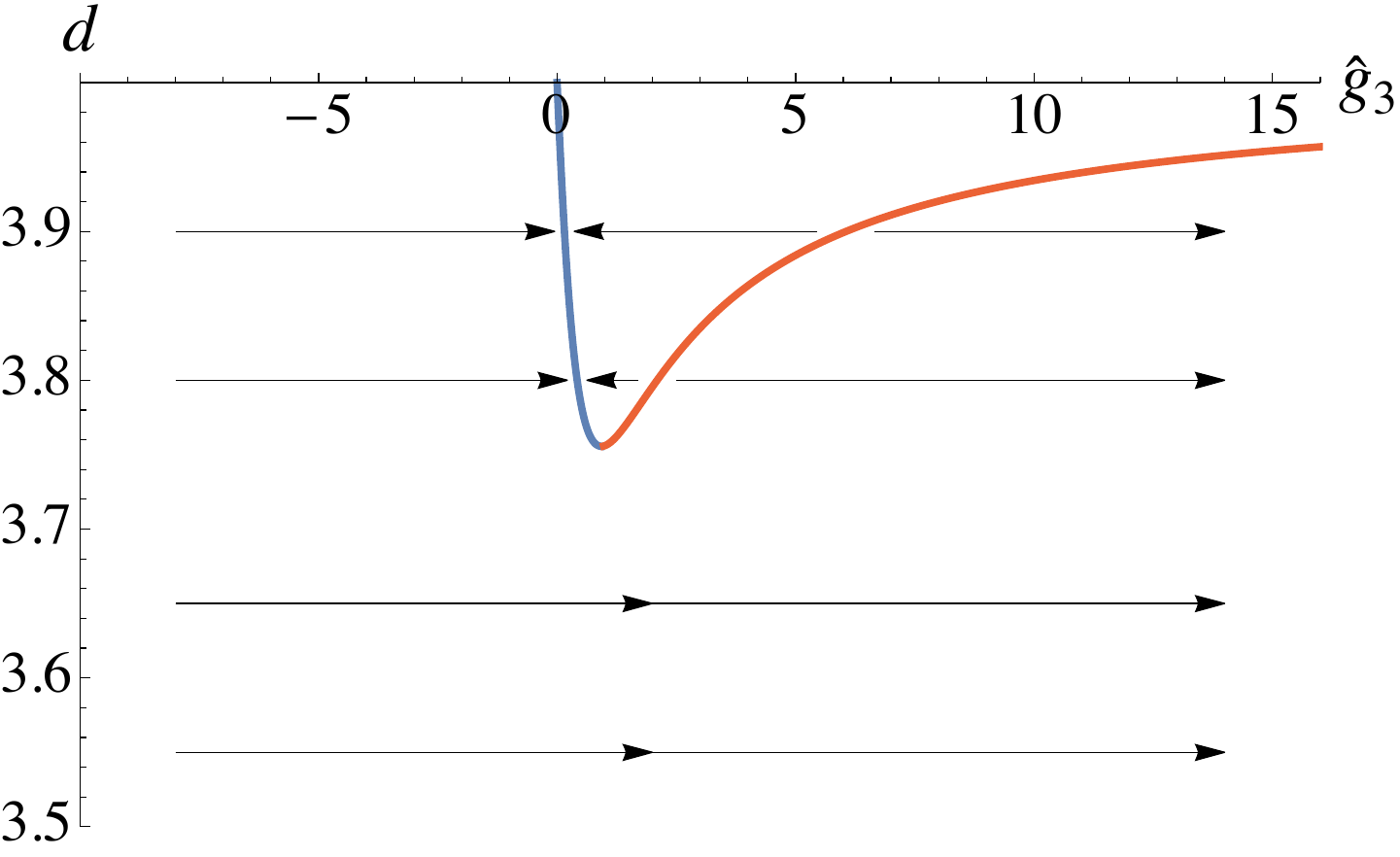}\\
\vspace{-2mm}$\bm\vdots\qquad~$\\
\includegraphics[width=0.95\columnwidth]{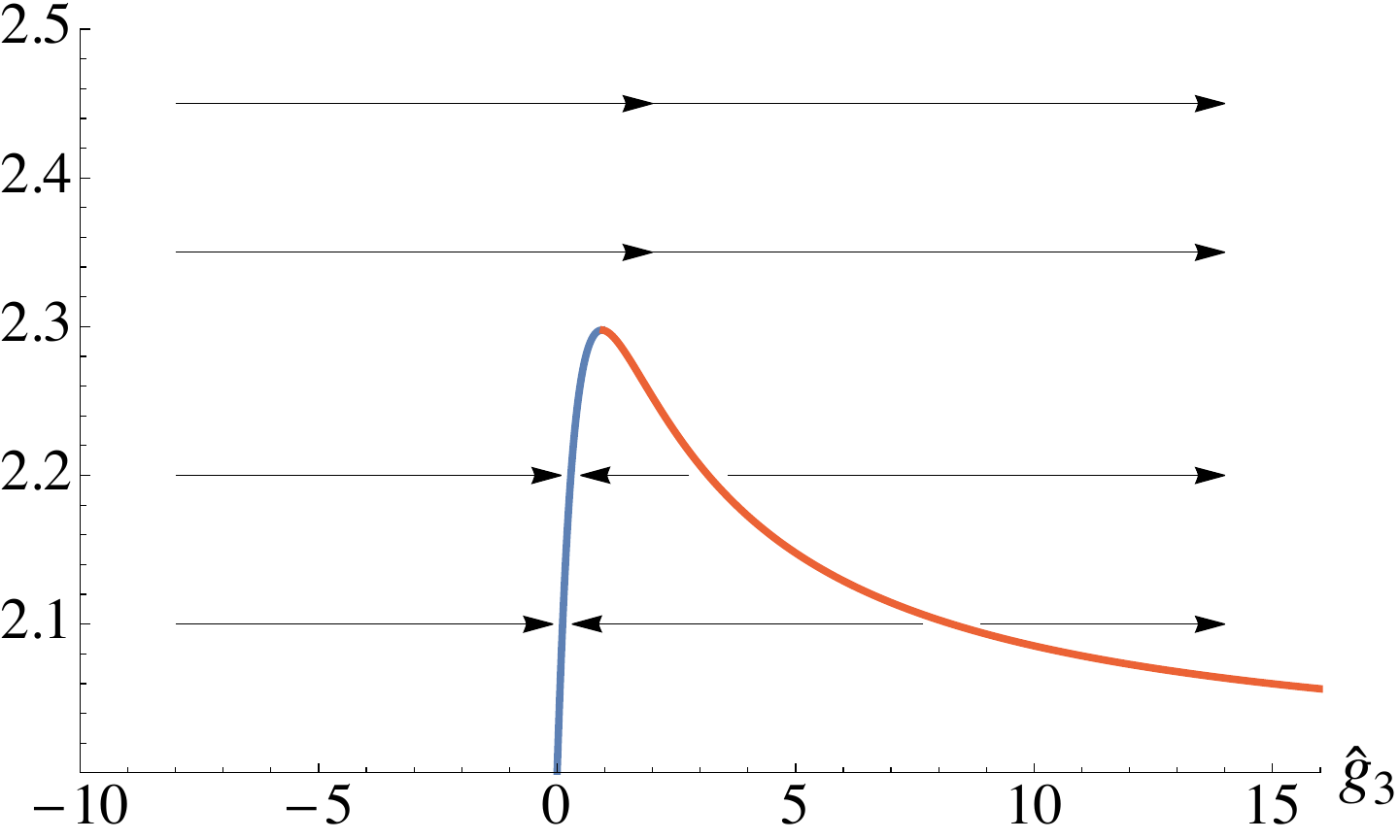}
\caption{\label{fig:flow}
RG flow of the dimensionless three-boson coupling $\hat{g}_3$ toward IR (arrow) at varying spatial dimensions $d$.
While a pair of IR (blue) and UV (red) fixed points exists for $2<d<2.30$ or $3.76<d<4$, they disappear for $2.30<d<3.76$.}
\end{figure}

We now determine the cutoff dependence of $\hat{g}_3$ by requiring the scattering amplitude to be cutoff independent at low frequency and wave vector.
This can be performed only numerically, but we find that our accurate numerical solution agrees well with the approximate analytical solution presented in Ref.~\cite{Mohapatra:2018} up to relative differences of 8\%.
The resulting RG flow of $\hat{g}_3$ is shown in Fig.~\ref{fig:flow}, where two distinct behaviors are found.
For $2<d<2.30$ or $3.76<d<4$~\cite{Nielsen:2001}, there exists a pair of IR and UV fixed points and the former corresponds to no three-boson interaction, while the latter to the binding transition of three bosons beyond which their bound state is formed.
However, the two fixed points collide with each other at $d=2.30$ or $3.76$ and then disappear into the complex plane for $2.30<d<3.76$.
In particular, the RG equation for $d=3$ can be obtained as
\begin{align}
\Lambda\frac{\d\hat{g}_3}{\d\Lambda}
= -\frac{1+s_0^2}{2}\left(\frac{\hat{g}_3^2}{c} + c\right)
+ \frac{1-s_0^2}{2}\,\hat{g}_3,
\end{align}
where $s_0\simeq1.00624$ and $c\simeq0.878658$ are numerical constants~\cite{Bedaque:1999a,Bedaque:1999b,Braaten:2011}.
Because the $\beta$ function is strictly negative, $\hat{g}_3$ keeps on growing toward IR even when it vanishes at UV.
Therefore, the scale invariance under $z=2$ as well as the nonrelativistic conformal invariance at the binding transition of two bosons is actually lost for $2.30<d<3.76$ as a consequence of the fixed-point annihilation~\cite{Kaplan:2009,Mohapatra:2018}.

Even more interestingly, the above RG equation for $d=3$ can be integrated, leading to a limit cycle solution of
\begin{align}
\hat{g}_3 = c\,\frac{1-s_0\tan(s_0\ln\Lambda/\Lambda_*)}{1+s_0\tan(s_0\ln\Lambda/\Lambda_*)},
\end{align}
where $\Lambda_*$ is an integration constant~\cite{Bedaque:1999a,Bedaque:1999b,Braaten:2011}.
Although $\hat{g}_3$ varies under a continuous change of $\Lambda$, it does not under a discrete change of $\Lambda\to e^{-\pi n/s_0}\Lambda$ ($n\in\Z$) so as to obey the discrete scale invariance.
In particular, an infinite sequence of divergences under RG in the three-boson coupling indicates an infinite sequence of three-boson binding energies, $E_n\propto e^{-2\pi n/s_0}\Lambda_*^2$, which is known as the Efimov effect~\cite{Efimov:1970,Efimov:1973}.
While such discrete scale invariance for three bosons persists for four bosons as well~\cite{Deltuva:2013}, whether it persists for an arbitrary number of bosons or not remains unestablished.

\section{Conclusion}\label{sec:conclusion}
We now know that the KPZ equation can be mapped onto quantum mechanics of attractive bosons with a contact interaction, where the roughening transition above two dimensions corresponds to the binding transition of two bosons.
On the other hand, we also know that such critical bosons exhibit the Efimov effect for $2.30<d<3.76$, where the three-boson coupling turns relevant under RG so as to break the scale invariance down to a discrete one as a consequence of the fixed-point annihilation.
What does this imply in turn for the roughening transition?
Because the full scale invariance under $z=2$ is lost in Eq.~(\ref{eq:boson}), the asymptotic scaling of Eq.~(\ref{eq:scaling}) with $\chi=0$ and $z=2$ is no longer expected.
Instead, depending on whether the discrete scale invariance persists for an arbitrary number of bosons or not, we predict that the KPZ roughening transition in three dimensions shows either the discrete scale invariance or no intrinsic scale invariance, as indicated in Fig.~\ref{fig:phase_diagram}.
Further descriptions of each possibility are as follows.

If the discrete scale invariance persists in Eq.~(\ref{eq:boson}) at the critical point of $g_2$, i.e., all higher-order terms denoted by dots are irrelevant, it should be reflected in the roughening transition as well.
Therefore, the asymptotic scaling of Eq.~(\ref{eq:scaling}) is to be replaced with
\begin{align}
\<[h(t,\r)-h(0,\0)]^2\> \sim F_{s_0}\!\left(\ln r\Lambda_*,\frac{t}{r^2}\right),
\end{align}
where $F_{s_0}(\ln r\Lambda_*,t/r^2)=F_{s_0}(\ln r\Lambda_*+\pi/s_0,t/r^2)$ is a periodic function of its first argument obeying the discrete scale invariance under $\r\to e^{\pi n/s_0}\r\simeq(22.7)^n\r$ and $t\to e^{2\pi n/s_0}t\simeq(515.)^nt$.
This is indeed a fascinating possibility, where the Efimov effect emerges at the KPZ roughening transition!

On the other hand, if the discrete scale invariance is broken by some relevant higher-order term in Eq.~(\ref{eq:boson}) at the critical point of $g_2$, there should be no asymptotic scaling intrinsic to the roughening transition.
Therefore, depending on from which side it is approached, the asymptotic scaling of Eq.~(\ref{eq:scaling}) is to be discontinuous at the roughening transition either with $\chi=-1/2$ and $z=2$ (smooth side) or with $\chi>0$ and $z=2-\chi$ (rough side).
This is a kind of first-order phase transition and may be considered as a nonequilibrium analog of the weak first-order phase transition as a consequence of the fixed-point annihilation~\cite{Gorbenko:2018a,Gorbenko:2018b}.

It is needless to say that our prediction should be tested with accurate numerical simulations toward the roughening transition in three dimensions~\cite{Yan:1990,Forrest:1990,Pellegrini:1990,Pellegrini:1991,Moser:1991,Tang:1992} or more rigorous mathematical approaches~\cite{Magnen:2018,Gu:2018,Comets:2020}.
Because the KPZ equation is equivalent to the stochastic Burgers equation and a directed polymer in a random medium~\cite{Kardar:1986}, our prediction equally applies to these problems, where the roughening transition corresponds to turbulent and pinning transitions, respectively.
Our observation of the relevant three-boson coupling, which is formally translated into $\eta^2$ added to Eq.~(\ref{eq:KPZ}), also raises a question about the well-definedness of the KPZ equation specific to three dimensions.
Finally, we conclude that the possible connection between the two seemingly distinct subjects, the KPZ equation in nonequilibrium statistical physics and the Efimov effect in quantum few-body physics, uncovered in this paper deserves further studies.

\acknowledgments
The authors thank Kazumasa A.\ Takeuchi for valuable discussions.
This work was supported by JSPS KAKENHI Grants No.\ JP17K14301 and No.\ JP18H05405.

\end{document}